\documentclass[12pt]{article}
\usepackage{amsmath, amsthm, amssymb, amsfonts}
\usepackage{hyperref}
\newtheorem{definition}{Definition}
\newtheorem{proposition}{Proposition}
\newtheorem{corollary}{Corollary}
\newtheorem{lemma}{Lemma}
\newtheorem{protocol}{Protocol}
\newcommand\pfun{\mathrel{\ooalign{\hfil$\mapstochar\mkern5mu$\hfil\cr$\to$\cr}}}

\DeclareMathOperator*{\concat}{\scalerel*{||}{\sum}}
\usepackage{scalerel}
\usepackage{stmaryrd}
\usepackage{enumitem}
\author{Alex Shafarenko}

\begin{document}

\title{Winternitz stack protocols}
\author{Alex Shafarenko}

\date{}
\maketitle
\begin{abstract}
This paper proposes and evaluates a new bipartite post-quantum digital signature protocol based on Winternitz chains and the HORS oracle. Mutually mistrustful Alice and Bob are able to agree and sign a series of documents in a way that makes it impossible (within the assumed security model) to repudiate their signatures. The number of signatures supported by a single public key is limited by a large number but the security of the signature scheme is not diminished by repeated application. A single public key supports both parties. Some ramifications are discussed, security parameters evaluated and an application area delineated for the proposed concept.
\end{abstract}

\section{Introduction}

Our focus is on supporting multivendor embedded devices that require guaranteed nonrepudiation. Such devices often occur in automotive, aerospace and other safety-critical applications, as well as in all kinds of medical technology. Digital signatures that utilise symmetric ciphers are based on sharing a confidential key. The same key is used for both signing messages and validating them, so the signer can always repudiate its signature by claiming that the document was actually signed by the validator themselves. Public-key cryptography does not quite solve this problem. First of all, it is vulnerable to quantum attacks and cannot be relied on in a future-proof technology. Secondly, and perhaps more importantly from a practical point of view, signature calculation and verification require a volume of computations that can be too large for an embedded device operating on a tight power budget.

We attempt to address both issues by proposing a hash-based signature. Such signatures exploit the one-way nature of a cryptographic hash to create an effective public-key/private-key pair. Large random integers are used as the private key, and their hashes are published to form the public key. When a new message requires a signature, the signer reveals certain integers of the private key. Their choice uniquely identifies the message, and since the only principal who can show the pre-image of a hash is the one who created that hash in the first place, a collection of pre-images reliably linked to a message can serve as its signature.

There are quite a few hash-based signature schemes that are well developed and understood, see Section \ref{sec:rel} for related work. We draw our inspiration from Winternitz's idea to apply a hash to a random seed repeatedly to create a chain, and Reyzin and Reyzin's idea of random index sets (multisets in our approach) generated by a hash and used to index an array of key-pairs. However, our proposal differs from the related work by the fact that we build a signature stack using a number of long Winternitz chains (we call this a Winternitz fabric), and push fixed size multi-set frames on it rather than variable-cardinality subsets, see Section \ref{sec:WSP}.  This way we achieve security that does not diminish under repeated application, so ours is not a {\em few-time} signature scheme, but actually a constant security scheme, albeit the capacity to sign is limited by the size of the fabric. Having said that, for a very modest storage capacity (single gigabytes), a million signatures can be accommodated without recalculation. Storage capacity can be traded in for hash recalculations and under realistic assumptions the storage requirements can be reduced to megabytes without significantly altering the amount of work required for signing, see section \ref{sec:prac}. We show that 256-bit security (128-bit post-quantum) is easily achieved by our scheme. The computational burden on the parties after the protocol launch amounts to a few tens of hash calculations by the verifier (same as the original HORS by Rayzin and Rayzin \cite{HORS}) and only one or two by the signer --- a negligible amount compared to the computations involved in a public-key signature scheme such as ECDSA.

The most unusual feature of our solution is its ability to support a mutual signature of two parties using a single public key by repurposing the confirmation pre-image for signing the verifier's approval, see Section \ref{sec:MAWS}. We propose an extreme version of this mutual protocol, which we call Reverse Winternitz Stack (RWS), where Alice only signs messages received from Bob, Section \ref{sec:RWS}. As a result Alice becomes Bob's signature server without requiring any trust between them. If Alice signs on Bob's behalf, Bob is always able to prove the signature false, but if the signature is genuine, then Bob is unable to repudiate. Alice has instant assurances from the protocol that Bob's signature is genuine, but any proof for a third party requires dumping Bob's stack. The stack contains digests of all documents that Bob's ever signed with Alice as well as additional protocol data the size of Alice's public key (hundreds of KB); this gives rise to a small communication requirement in the order of 1Mbyte. In most scenarios that involve a guarantee of nonrepudiation, a third-party proof is only required after a major event (car breakdown, aircraft malfunction, etc.) to adjudicate on the cause in a multivendor environment.

Another remarkable feature of RWS is its communication asymmetry. Bob receives on the order of 1Kb of data from Alice, but his transmission needs are limited to only 64 bytes per signature, same as it would be for the 256-bit ECDSA. In an IoT situation, where messages are communicated over long distances via a low bit-rate radio, {\em transmission} requires much power to radiate a strong enough signal to reach the network hub, while {\em reception} involves only digital signal processing. Also the maximum radiated power and the transmitter duty cycle of an IoT radio is limited by law to enable public use without harmful interference. By contrast, the IoT network hub is allowed to have a more powerful transmitter (up to a factor of 10), higher duty cycle (again a factor of 10) and an elevated full-size antenna to be able to transmit much greater volumes of data. The RWS protocol nicely matches this asymmetry. 

To end this section we summarise the contributions of the paper:
\begin{itemize}
\item The idea of Winternitz stack and an analysis of its security properties 
\item Three signature protocols based on a Winternitz stack with undiminished security for a large but limited number of signatures 
\item Analysis of the protocol resource footprints
\end{itemize}

The next two sections present the basic principles, notations and some security properties of the Winternitz stack. Sections \ref{sec:WSP},\ref{sec:MAWS} and \ref{sec:RWS} describe the signature protocols. Section \ref{sec:app} discusses possible applications, Section \ref{sec:rel} presents related work and finally there are some conclusion.

\section{Principles of Winternitz stack signature}
\begin{figure}
\begin{center}
\includegraphics[width=0.5\textwidth]{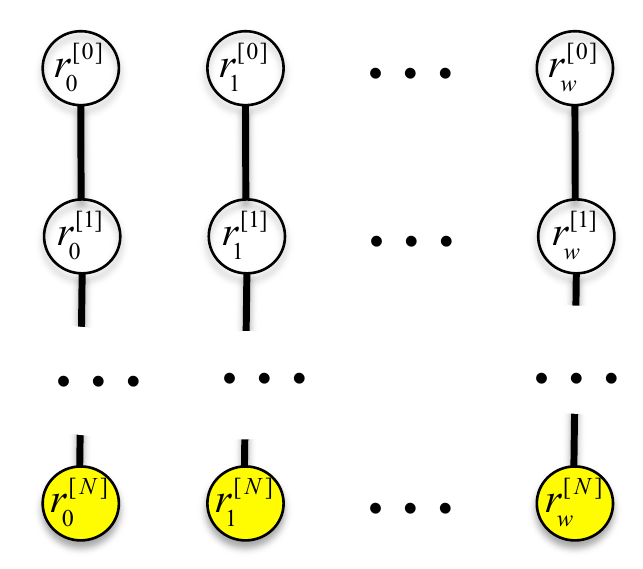}
\end{center}
\caption{Winternitz fabric. The shaded nodes represent the fabric edge. A vertical line followed down connects a value with its hash image\label{fig:fabric}}
\end{figure}

We begin with the standard definition of Winternitz chain:
\begin{definition}
For a cryptographic hash-function $H(x)$ and some arbitrary $r_0$ of the same bit-length as the image of $H$, the sequence \[
r_{i+1}=H(r_i),\; \hbox{for}\;\;0\le i<N
\]
is called a length-$N$ {\em Winternitz chain}. 
\end{definition}
\noindent Let us bring several chains together.
\begin{definition}[Winternitz fabric]
An indexed family of length-$N$ Winternitz chains $r^{[k]}_i$, where $k$ is the index, $0\le k<w$, $w$ is a power of 2, and $0\le i<N$, is called a {\em Winternitz $(w,N)$-fabric}, or just fabric for short. The constant $w$ is called the {\em width} and $N$, the length of the fabric. The indexed family $(E^{[k]})$,  where $E_k=r^{[k]}_N$ where $0\le k<w$ is called the {\em edge} of the fabric. When referring to the edge as a whole we will omit the index: $E$. 
\end{definition}
\noindent
See Figure \ref{fig:fabric} for an illustration. The width of a fabric is constrained to a power of 2, but the length of the chain is arbitrary. For practical purposes chain lengths of the order of one million or less should satisfy most demands for a digital signature. Notice that due to the hardness of the second preimage problem, at most one 
$(w,N)$-fabric can be produced given an arbitrary width-$w$ edge $E$ for any practical $N$.

Consider a length-$d$ sequence of binary strings $D_j$, $0\le j<d$, which we will call {\em documents}. Let a function 
\[
\Omega: {\mathcal B}\to {\mathcal M}_{\kappa,w}
\] 
be a random oracle, where ${\mathcal B}$ is a set of arbitrary-length binary strings and $\mathcal M_{\kappa,w}$ is a set of all cardinality-$\kappa$ multisets of integers taken from the range $\llbracket0,w-1\rrbracket$. Assume the oracle is such that the presence of those integers in the multiset is uncorrelated. A multiset $M\in{\mathcal M}_{\kappa,w}$ of this kind can be defined using its characteristic function $\chi_M:\llbracket0,w-1\rrbracket\to\llbracket0,\kappa-1\rrbracket$, such that for any $x\in\llbracket0,w-1\rrbracket$ $x$ occurs in $M$ $\chi_M(x)$ times. For convenience, define $\omega: {\mathcal B}\times \llbracket0,w-1\rrbracket\to\llbracket0,\kappa-1\rrbracket$ to be a function such that $M=\Omega(b)$ iff $\omega(b,k)=\chi_M(k)$ for any $b\in {\mathcal B}$ and $0\le k<w$. Clearly, for any $b\in {\mathcal B}$, 
\begin{equation}
\sum_{k=0}^{w-1}\omega(b,k)=\kappa\,.\label{eq:norm}
\end{equation}
Cardinality $\kappa$ is a security parameter. We assume it is fixed and will discuss the choice of it later on. In the sequel we will not show the dependency of any variables of interest on $\kappa$ explicitly. 

\begin{definition}\label{def:stack}
For an integer $d>0$, a depth-$d$ signature stack over a Wintenitz $(w,N)$-fabric $(r^{[k]}_i)$is a pair of indexed families $(D,T)$, where  
$D=(D_j)_{j\in\llbracket0,d-1\rrbracket}$ is a sequence of documents, and $T=(r^{[k]}_{N-\sigma(k)})_{k\in\llbracket0,w-1\rrbracket}$ is the top of the stack, with the function $\sigma:\llbracket0,w-1\rrbracket\to\llbracket0,N-1\rrbracket$ defined thus:\[
\sigma(k)=\sum_{m=0}^{d-1}\omega(\concat_{j=0}^m \hskip-0.5em D_j,k)\,,
\]
provided that $\sigma(k)\le N$ for all $k$ in its domain. (The double vertical bar denotes bit-string concatenation.)

A depth-0 signature stack is the pair $(\emptyset,(r^{[k]}_N)_{k\in\llbracket0,w-1\rrbracket})$  
\end{definition}
\begin{figure}
\begin{center}
\includegraphics[width=0.5\textwidth]{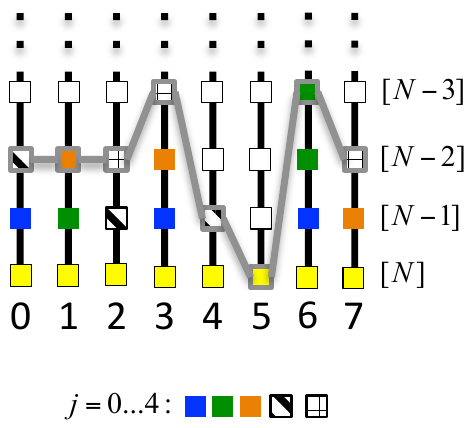}
\end{center}
\caption{An example of a depth-5 signature stack over a Winternitz $(8,N)$-fabric using a random oracle with $\kappa=3$. Empty boxes represent unused members of the fabric. The top of the stack $T$ is highlighted in grey and the fabric edge in yellow. The document family is not shown. \label{fig:stack}}
\end{figure}

\begin{corollary}\label{cor:mass}
For any depth-$d$ signature stack $(D,T)$ over a Wintenitz $(w,N)$-fabric,
\[
\sum_{k=0}^{w-1}\sigma(k) = d\kappa
\]
\end{corollary}
\begin{proof}
Sum in $k$ both sides of the equation for $\sigma$ in defition \ref{def:stack}, make the summation in $k$ innermost in the right-hand side and use Eq \ref{eq:norm}.
\end{proof}
\subsection{Fabric capacity} 
Typically what is signed is not the actual content but its cryptographic digest, so $D_j$ are usually hash-images of the actual documents to be signed. Also, to prevent a replay attack, the original content typically contains a random nonce. Under such assumptions the family $(D_j)$ is a collection of random values which makes the oracle output not only random but also, with a probability very close to 1, free from repetitions. Let us evaluate the capacity of the fabric to carry a signature stack of a large depth $d$. 

First visualise the fabric as $w$ vertical rods that balls can be slid unto, see Figure \ref{fig:stack}. Each document $D_j$ causes balls to be slid on some of the rods according to $\omega$, the total number of balls being $\kappa$. Their distribution over the rods is random and uncorrelated. In particular, it is possible but not very probable that some balls for a given document will be slid on the same rod. The distribution $\sigma(k)$ is the result of repeating ''the sliding of balls'' $d$ times using a total of $d\kappa$ balls. Choose one rod at random and observe that the probability for a ball to end up taking that rod is $1/w$. The number of balls on the rod after $d\kappa$ balls have been randomly distributed between the rods is governed by the binomial distribution, which gives us the obvious expectation $E=d\kappa/w$ and the standard deviation \[
\Delta=\sqrt{{d\kappa \over w}\left(1-{1\over w}\right)}\approx \sqrt{d\kappa \over w}\,,
\]
for large $d$. By Central Limit Theorem of statistics the distribution becomes close to Gaussian at large $d$. The rule of thumb is that fluctuations of a Gaussian random value very rarely exceed $6\Delta$ and so we calculate, for large $N$ that we are interested in, \[
{d_{{\rm max}}\kappa \over w}\lesssim N-6\sqrt{N}\,.
\]
The exact point at which the fabric will prove too small to support the signature stack over it depends on the documents $(D_j)$, but since each document only slides $\kappa$ balls on the rods, and since we will always use wide fabrics ($\kappa\ll w$), the process can be stopped very close to that point. For estimates we should neglect $\sqrt{N}$ compared to $N$ and use
\begin{equation}
d_{{\rm max}}\sim{wN\over\kappa}\,.\label{eq:capacity}
\end{equation}

\subsection{Security of the oracle} Next let us explore security properties of a signature stack. The lynchpin of security here is the fact that a random oracle $\Omega(D)$ makes the problem of a second preimage unfeasibly hard, provided that its codomain is a large set. An attacker trying to find some $D^\prime$ that has the same image as $D$, $\Omega(D^\prime)=\Omega(D)$ will have to make a number of attempts commensurate with $|{\mathcal M}_{\kappa,w}|$, which could be astronomically large. We find from elementary combinatorics that for a fabric of width $w$, an oracle with the security parameter $\kappa$ will yield one out of a possible \[
G=\binom{w+\kappa-1}{\kappa}
\]  
multisets of cardinality $\kappa$. For practical reasons, which we will explain later, we are interested in a small $\kappa$ of the order ten, while we are willing to consider fabrics of a width of a few thousand. Expanding the above for $\kappa\ll w$ and keeping the first nonvanishing term in $1/w$ we obtain
\begin{equation}
G={(w+\kappa-1)\times\ldots\times w\over \kappa!}\sim
{w^\kappa\over\kappa!}\left(1+{\kappa(\kappa-1)\over2w}\right)\,,\label{eq:oracle}
\end{equation}
Note that this is an estimate from below as all higher-order terms are positive. We are interested in the entropy of the oracle, which equals the binary logarithm of $G$. To get some idea how large it can be for 
$\kappa^2\lesssim w$ we neglect the term in brackets and use Stirling's formula for the factorial, which is accurate within 1\% even for $\kappa$ as small as 10:
\[
\log_2 G \approx \kappa\log_2(we/\kappa)-{1\over2}\log_2(2\pi\kappa)\,.
\]
Consider a fabric of width $w=4096$ and an oracle with $\kappa=31$ and find that $\log_2 G$ exceeds 259. Notice that it is marginally better than the security of SHA-256 with respect to a second preimage attack. 

A true random oracle is not feasible, but it can be approximated very well using the trick invented by Reyzin and Reyzin as they proposed their HORS scheme \cite{HORS}. Specifically we take a standard hash of the argument $D$ and pare it down to $\kappa\log_2w$ bits\footnote{remember that the width of a fabric is required to be a power of 2 by the above definition}. In our numerical example this would be $31\times12=372$ bits. A hash of this length is easy to compute by application of SHA-512, taking the first 372 bits of the result. All that remains is to partition the bit string into 12-bit chunks, interpret them as unsigned binary integers,  and to collect them into a multiset, which will represent the value of $\Omega(D)$. We call this implementation a {\em HORS oracle}.

Strictly speaking, a HORS oracle is not a proper random oracle even if we disregard the difference between a cryptographic hash and a genuine random function. The random oracle makes a random selection of a multiset from the domain ${\mathcal M}_{\kappa,w}$; any $B\in{\mathcal M}_{\kappa,w}$ is selected with the same probability. The HORS oracle (again, ignoring the non-random nature of the hash) does not select a multiset; it selects $\kappa$ random indices from the interval $\llbracket0,w-1\rrbracket$, which may or may not be  pairwise distinct. If they are, then the multiset is a set of cardinality $\kappa$ and its statistical weight in the hash function domain is $\kappa!$. If the index set has one binary collision, its statistical weight is only one half of that (accounting for the transposition of the two identical index values that does not change the multiset). Further collisions degrade the statistical weight even further. Consequently, the entropy of the HORS oracle may well be significantly different from the estimate given by Eq \ref{eq:oracle}. How much different? 

It should be noted that the number of proper multisets (i.e. multisets that are not sets) is small compared to the number of sets in the oracle's domain\footnote{Also see \cite{binomials} for many useful asymptotic bounds for binomial coefficients}: 
\[
G_{+} = \binom{w+\kappa-1}{\kappa} - \binom{w}{\kappa} 
\sim {w^\kappa\over\kappa!}{\kappa(\kappa-1)\over w}\sim R\gamma\,,
\]
where we kept the first nonvanishing term in the expansion in the birthday\footnote{so named as it defines the probability of collision in the birthday paradox} parameter $\gamma=\kappa^2/w$. The factor $R=w^\kappa/\kappa!$ corresponds to the number of distinct sets (ignoring the collisions) swept by the indices as they independently cover their value interval $\llbracket0,w-1\rrbracket$. The most frequent proper multiset has one binary collision. There are \[
G_1=w\binom{w-1}{\kappa-2}\sim \gamma\binom{w}{k}\sim \gamma R
\]
of those, again keeping to the first nonvanishing term in $\gamma$. Comparing this with the expression for $G_{+}$ above, we find that the contribution of multiple collisions is higher order in $\gamma$. Indeed the next term corresponds to {\em two binary} collisions since there are a factor of $\kappa/w$ fewer multisets with one tertiary collision. The statistical weight of the former is  
\[
\binom{w}{2}\binom{w-2}{\kappa-4}\sim {1\over2}\gamma^2R\,,
\]
and can safely be neglected for small enough $\gamma$ (in our example $\gamma\approx1/4$) along with the rest of the higher-order terms. By contrast, the number of multisets that are sets can be approximated to the first order in $\gamma$ as 
\[
G_0=\binom{w}{\kappa}=R\left(1-{\gamma\over2}\right)\,.
\]
We can now construct an approximate probability distribution function (PDF) for multisets by assuming that a mutiset is either a set or it has one binary collision:
\[
f(s)=\begin{cases}1/z&\hbox{if $s$ is a set,}\\ 1/(2z)&\hbox{otherwise,} \end{cases}
\]
where $z$ is the normalising constant that satisfies the following:
\[
G_0{1\over z}+G_1{1\over2z}=1\,,
\]
which gives us $z=R$. Finally, summing over all sets and proper multisets and keeping to the main order in $\gamma$ we arrive at the entropy value
\[
H=-G_0{1\over R}\log_2{1\over R}-G_1{1\over 2R}\log_2{1\over 2R}=\log_2 R +{\gamma\over2}\sim \log_2 G -
{\log_2 e - 1\over 2}\gamma\,,
\]
where the last step is achieved by taking the binary logarithm of Eq \ref{eq:oracle} and keeping to the first order in $\gamma$. Informally, proper multisets expand the alphabet of the HORS oracle output compared to just proper sets, thus increasing its entropy, however they make the PDF uneven and this reduces the entropy by almost the same amount. As a result, a HORS oracle with a moderately small $\gamma$ has almost exactly the same entropy as the ideal oracle.  

\subsection{Stack security} 
Security properties of signature stacks are due to the following two propositions. 
\begin{proposition}\label{prop:fit}
For any depth-$d$ signature stack $(D,T)$ over a fabric, it is computationally hard to find a family $D^\prime\ne D$ such that $(D^\prime,T)$ is a depth-$d$ signature stack over the same fabric.  
\end{proposition}
\begin{proof}
Essentially the attacker would have to solve the following set of simultaneous equations for $(D^\prime_j)$:
\[
\sum_{m=0}^{d-1}\omega(\concat_{j=0}^m \hskip-0.5em D^\prime_j,k) = \sum_{m=0}^{d-1}\omega(\concat_{j=0}^m \hskip-0.5em D_j,k)\;\;, 0\le k<w
\]
The left-hand side is the sum of $d$ terms. Changing any document $D_{l}$ in any way at all would result in a near-random unpredictable change of all terms of the sum for which $l\le m<d-1$. The oracle is not invertible, so the only way to find a preimage is to brute-force the whole domain, i.e. the combined length of the affected documents after freezing the documents with indices less than $l$. This may not even be possible if \[
\sum_{m=0}^{l-1}\omega(\concat_{j=0}^m \hskip-0.5em D^\prime_j,k) > \sum_{m=0}^{d-1}\omega(\concat_{j=0}^m \hskip-0.5em D_j,k)
\]
for some $k$, $0\le k<w$, but there is always an option not to change documents with indices less than $l$ (and that guarantees that the above inequality does not hold.

Clearly the least work the attacker needs to do is when $l=d-1$, i.e. only the last document is brute forced, but then the attack is equivalent to a second preimage attack, since the required value of the function $\sigma$ is fixed by the right-hand side. We conclude that an alternative depth-$d$ signature stack with the same fabric and top is computationally unfeasible.
\end{proof}

\begin{corollary} 
Given $d$, the fabric edge $E$ and the stack top $T$, at most one family of documents can be found that makes $(D,T)$ a signature stack over the fabric defined by $E$. 
\end{corollary}
Notice that $T$ can be validated without knowledge of the fabric values by attempting to compute $\beta(T_k,E_k)$
where $\beta(x,y)$ is the least positive integer $i$ such that 
\[
\underbrace{H(H(\ldots H}_{i\ \text{times}}(x)\ldots))=y\,.
\]
If $x=y$, we define $\beta(x,y)=0$.

Now $T$ can only be valid if the following equation is well defined and holds:
\begin{equation}\label{eq:mass}
\sum_{k=0}^{w-1} \beta(T_k,E_k)=d\kappa\,.
\end{equation}
It should be noted $\beta$ is a partial function ${\cal B}\times{\cal B}\pfun{\mathbb N}$, since the value of $i$ that satisfies its definition may not exist. However, if the maximum length of a fabric is limited (and it has to be, since fabrics have to be computed fully {\em before} use), the knowledge of the maximum possible fabric length $N_{\rm max}$ is sufficient to stop the computation and render the above equation undefined.

Since $T_k=r^{[k]}_{N-\sigma(k)}$, $\beta(T_k,E_k)=\sigma(k)$, if defined. Given the documents $D$, $\sigma$ uniquely defines the stack $(D,T)$ over the fabric with the edge $E$, provided that the fabric length is at least $1+\max_k\sigma(k)$. We will use $(D,\sigma)$ and $(D,T)$ interchangeably where it does not create a confusion.

\begin{definition}[Substack]
For a depth-$d$ stack $S=(D,\sigma)$ over a $(w,N)$-fabric, a depth-$d^\prime$ stack $(S^\prime=(D^\prime,\sigma^\prime)$ over the same fabric is a substack of $S$ if for all $0\le k<w$, $\sigma^\prime(k)\le \sigma(k)$.
\end{definition}

\begin{proposition}\label{prop:mass}
Consider a depth-$d^\prime$ substack $S^\prime=(D^\prime,\sigma^\prime)$ of a depth-$d$ stack $S=(D,\sigma)$. If $d^\prime=d$, then $\sigma^\prime=\sigma$.
\end{proposition}
\begin{proof}
Proof by contradiction. Assume $\sigma^\prime\ne \sigma$, and since for all $0\le k<w$, $\sigma^\prime(k)\le \sigma(k)$, then $(\exists k_0) \sigma^\prime(k_0)<\sigma(k_0)$. But then 
\[
\sum_{k=0}^{d^\prime-1}\sigma^\prime(k) < \sum_{k=0}^{d-1}\sigma(k)\,,
\]  
By Corollary \ref{cor:mass} we have\[
\kappa d^\prime < \ \kappa d\,,
\]
which is a contradiction since $d^\prime=d$.
\end{proof}
Notice that Proposition \ref{prop:mass} does not generalise down. If $d^\prime<d$, especially when $d^\prime\ll d$, almost any documents $(D_j)$ will place the top of the substack  lower on the fabric than the larger stack's top. Indeed to make the substack tall with a small $d^\prime$ would require a document whose distribution $\omega$ over the fabric  is restricted to very few values of $k$ which is very improbable for a random oracle or a high-quality emulation of it. Consequently there exists plenty of substacks of a given stack with a smaller depth.

\begin{lemma}[Stack security]\label{lem:stack-sec}
If $T$ is known to be the top of a depth-$d$ stack $(D,T)$ over a {\em private} fabric $F$ with a public edge $E=E(F)$, the pair $(E,T)$ is sufficient to find $d$ and uniquely identify $D$. It is computationally hard for an adversary with no knowledge of the rest of $F$ to produce an alternative $D^\prime\ne D$, such that $(D^\prime,T^\prime)$ with any $T^\prime$ is a valid depth-$d$ stack over a fabric with the same edge $E(F)$.    
\end{lemma}
\begin{proof}
To find $d$ from $T$, use Eq \ref{eq:mass}. Next, observe that knowledge of $(D,T)$ is always sufficient to reconstruct all members of the fabric that the stack and any of its substacks occupy down to the edge $E$. Although given just that knowledge it is possible to construct a valid substack $(D^\prime\ne D,T^\prime)$ for some arbitrary documents $(D^\prime_j)$, but according to Proposition \ref{prop:mass} the substack would have to be of a depth less than $d$, which contradicts the premise of the Lemma. If $T^\prime=T$, changing even a single $D_j$ to $D^\prime_j\ne D_j$ without changing $T$ is computationally hard according to Proposition \ref{prop:fit}. Finally, if $(D^\prime,T^\prime)$ is not a substack of $(D,T)$, to produce $T^\prime$ one would require fabric elements that cannot be derived from $T$, and the premise states that the fabric is private. 
\end{proof}

\section{Ancillary operations}

\subsection{Stack push} 

Signature stacks over a fabric are inherently sequential. It is possible to extend a stack to accommodate an extra document provided that the fabric is accessible.  

\begin{definition}\label{def:push}
Stack push $p$ is a partial function ${\bf p}:{\mathcal D}\times{\mathcal S}_d\pfun {\mathcal S}_{d+1}$, where ${\mathcal D}$ is, as before, a set of all finite binary strings and ${\mathcal S}_d$ is an indexed family of sets of all depth-$d$ stacks over some fixed $(w,N)$-fabric $(r^{[i]}_k)$\/{\rm :} 
\[
{\rm\bf p}(\delta,((D_j)_{0\le j <d},(r^{[k]}_{N-\sigma(k)})_{0\le k <w}))\triangleq((D^\prime_j)_{0\le j <d+1}, (T^\prime_k)_{0\le k <w})\,,
\] 
where 
\[
D^\prime_j=D_j\;\hbox{\em for}\; 0\le j<d\,, 
\]
\[
D^\prime_d=\delta\,,
\]
\[
\sigma^\prime(k)=\sigma(k) + \omega(\concat_{j=0}^d \hskip-0.5em D^\prime_j,k)\;\hbox{\rm for }\;0\le k <w\,,
\]
and
\[
T^\prime=\left(r^{[k]}_{N-\sigma^\prime(k)}\right)_{k\in\llbracket0,w-1\rrbracket}\hbox{\rm for }\;0\le k <w\,,
\]
provided that all such $r$ exist in the fabric. Otherwise the result is undefined.
\end{definition}
Observe that\footnote{ignoring the infinitesimal probability of the random oracle producing all-zeros (more than 300 zeros in our numerical example)} at least for some $k$, $\sigma^\prime(k)>\sigma(k)$. This means that the stack push always depends on unused members of the fabric and requires access to it.

\subsection{Operations on indexed families} 
We require two ancillary operations on indexed families which we will denote as similar operations on sets.
\begin{definition}\label{def:famdif}
Let $A$ and $B$ to be two indexed families $A=(A_i)_{i\in C}$ and $B=(B_i)_{i\in C}$ with indices from the same finite $C\subset{\mathbb Z}^{+}$. We define the {\bf difference} $A-B$ as the indexed family $A-B=(A_i)_{i\in C^*}$, where 
\[
C^*=\{i\in C\mid A_i\ne B_i\}
\]
\end{definition}
\begin{definition}\label{def:famsum}
Let $A$ and $B$ to be indexed families $A=(A_i)_{i\in C}$ and $B=(B_i)_{i\in C^*}$, where $C\subset{\mathbb Z}^{+}$ is some finite set and $C^*\subseteq C$. We define the {\bf sum} $A+B$ as the indexed family $(Q_i)_{i\in C}$, where 
\[
Q_i = 
\begin{cases}
B_i&\hbox{if}\;\; i\in C^*\\
A_i&\hbox{otherwise}
\end{cases}
\]  
\end{definition}

\begin{proposition}
For any two finite indexed families $A=(A_i)_{i\in C}$ and $(B_i)_{i\in C}$, where C is a finite index set, \[
B+(A-B) = A
\]
\end{proposition}
\begin{proof}
(by cases) For a given index value $i$, if $A_i\ne B_i$ then, by Definition \ref{def:famdif}, $(A-B)_i=A_i$ and $i$ belongs to the index set of $A-B$. But if it does, then by Definition \ref{def:famsum}
\[
(B+(A-B))_i=(A-B)_i = A_i\,.
\] 
Otherwise $B_i=A_i$ and, by Definition \ref{def:famdif}, the index value $i$ does not belong to the index set of $A-B$. Then by Definition \ref{def:famsum}
\[
(B+(A-B))_i=B_i=A_i\,.
\] 
\end{proof}
If two parties share a family of bit-strings $B$ and at some point one needs to send to the other a similar family $A$ which has many members common with $B$, it would be sufficient to communicate $A-B$, which has a much smaller index set, and then the receiving party will restore $A$ by computing $B+(A-B)$.

\begin{definition}
Given an indexed family $X=(X_i)_{i\in I}$, where $I$ is a finite index set $I\subset{\mathbb Z}^{+}$, and some element $x$, the {\bf extension} of $X$ with $x$ is the family $X^\prime = X\triangleright x$ such that $X^\prime=(X^\prime_i)_{i\in I^\prime}$, 
\[I^\prime = I \cup \{\max_I i+1\}\] 
and \[
X^\prime_i=
\begin{cases}
X_i&\hbox{if}\;\; i\in I\\
x&\hbox{otherwise}
\end{cases}
\]
\end{definition}

\subsection{Validator}

The following definition introduces a validity test when a stack is extended with a new document $\delta$.

\begin{definition}\label{def:check}
Consider a stack $S=(D,T)$, $D=(D_j)_{j\in\llbracket0,d-1\rrbracket}$ and $T=(T_k)_{k\in\llbracket0,w-1\rrbracket}$, over a width-$w$ fabric of sufficient length,
a document $\delta$, and an indexed family of binary strings $(\tau_k)_{k\in \Gamma}$, with some $\Gamma\subset\llbracket0,w-1\rrbracket$ such that $|\Gamma|\le\kappa$. We define the validator predicate $\epsilon(\delta,\tau,S)$ to be true iff for all $k\in\llbracket0,d-1\rrbracket$, $\beta(T^\prime_k,T_k)$ exists and
\[
\omega(\concat_{j=0}^d \hskip-0.5em D^\prime_j \concat \delta,k)=\beta(T^\prime_k,T_k),
\]
where the indexed family $T^\prime=T+\tau$.
\end{definition}

\begin{proposition}
Given some $S=(D,T)$, $\delta$ and $\tau$ as per Definition \ref{def:check}, if $\epsilon(\delta,\tau,S)$ then $(D\triangleright\delta,T+\tau)$
is a valid depth-$(d+1)$ stack over the same fabric.
\end{proposition}
\begin{proof}
follows from Definition \ref{def:stack}
\end{proof}

\section{Winternitz Stack Protocol\label{sec:WSP}}

The structures presented so far can be used to create a bipartite protocol where {\em neither} party can repudiate a transaction. Under public-key cryptography, the signer is unable to repudiate a properly signed document since the verifier holds the signer's authenticated public key and can prove to an adjudicator that whoever signed the document had to have knowledge of the signer's private key. The parties are assumed to be mutually adversarial to exclude collusion\footnote{Note that no bipartite protocol can prevent collusion since the parties can destroy any documents and signatures by mutual consent and run the protocol from the beginning.}.  

\paragraph{Channel model.} We wish to minimise assumptions about the communication channel between the parties, bearing in mind that beneficial channel properties may depend on trust and/or shared confidential information. We require weak integrity, i.e. that a message sent by one party to the other and which fails the other party's validation test will be received intact after a finite maximum number of re-transmissions that does not depend on the message. No authentication of communicating parties is required; however countermeasures must be put in place to prevent an adversary from injecting messages in the channel at a rate that overwhelms the bona fide recipient. Because both parties are interested in progress, they can share a weak secret based on which all messages are extended with a MAC. However, even if no secret is shared, all protocols we present in the sequel require little computation at the receiving end for message validation (at most $\kappa+1$ hash evaluations per message), so in practice each protocol contains its own DoS countermeasure, which may or may not be combined with other defences depending on the threat model. We mark received values with an asterisk $*$ to emphasise that they are not necessarily the same as those sent.
Consider the following  

\begin{protocol} [Bipartite Winternitz Stack (BWS) protocol]\label{prot:bipartite}
~\\
Parties: Alice(signer) and Bob(verifier)\\
Protocol parameter: fabric width $w\in\mathbb{N}$, $w$ is a power of 2.\\
{\bf Initially:}
\begin{enumerate}[label={\bf I}\rm\arabic*]
\item In private: Alice chooses $N$, produces a random $(w,N)$-fabric $F$ and saves it in local secure storage. 
\item \label{step:public-key} Alice publishes the fabric edge $E=E(F)$ and authenticates it out of band. $E$ is now Alice's {\em public key}
\item Alice invites Bob to participate in up to $L$ transactions, $L\lesssim wN/ \kappa$, see Eq \ref{eq:capacity}. The invitation and the value of $L$ need not be authenticated\/\footnote{Here and until further notice: we ignore DoS issues.}.  
\item Bob creates a random length-$L$ Winternitz chain $q^{[k]}$ and authenticates $Q=q^{[L-1]}$ out of band. $Q$ is now Bob's public key, good for $L$ transactions. Also Bob produces the initial stack $S_0=(\emptyset,E)$ and saves it in local storage.
\item Alice prepares document $\delta=L\concat Q$ and computes a depth-1 stack over $F$: $S_1 = {\rm\bf p}(Q,(\emptyset,E))=(D_1,T_1)$, stores it, then communicates $\tau=T_1-E$ to Bob.
\item Bob prepares the same $\delta$ and checks $\epsilon(D_1,\tau,S_0)$. If true, Bob sends $q^{[L-2]}$ back to Alice and saves\footnote{\label{foot:stackstore}only the latest stack needs to be kept in storage as all the previous ones are its substacks.} $S_1=(\emptyset\triangleright D_1,E+\tau)$ in local storage. If false, Bob ignores $\tau$ and waits for retransmission.  
\end{enumerate}
{\rm (Any non-receipt of a protocol message so far can be overcome by Automatic Repeat Query (ARQ) safely. If ARQ fails, this constitutes denial of service by the non-responding party or a protocol violation by the sender. Either way, the protocol fails.)}

\vskip1em
\noindent
{\bf Repeat for $j=2..L-1$:}

\begin{enumerate}[label={\bf R}\rm\arabic*]
\item \label{step:alice} When Alice wishes to sign the next document $\delta_j$, she computes \[
S_{j}={\rm\bf p}(\delta_j,S_{j-1})=(D_j,T_j)\,,
\]
\noindent stores $S_j$ and sends to Bob $\delta_j$ and $\tau_j=T_j-T_{j-1}$.

\item \label{step:bob}Bob receives ($\delta^*$,$\tau^*$). Bob checks if he received and validated $\delta^*$ from the previous round $j-1$, and if so, (re-)sends $q^{[L-j-2]}$ to Alice\footnote{here, as before, we assume that any document that Alice signs contains a nonce making it impossible for two documents to be the same.} and remains in step \ref{step:bob} of the current round. 

Otherwise, $\delta^*=\delta^*_j$ is fresh for the current round. Bob computes  $\epsilon(\delta^*_j,\tau^*_j,S_{j-1})$. {\bf If true}, Bob stores\footnote{see footnote \ref{foot:stackstore}} $S_j=(D_{j-1}\triangleright\delta^*_j,T_{j-1}+\tau_j)$, where $(D_{j-1},T_{j-1})=S_{j-1}$, and sends $q^{[L-j-1]}$ to Alice as the acknowledgement. The round is finished. {\bf If false}, Bob sends a NAK and ignores $\delta^*_j$ and $T^*_j$ as if they had not been received\footnote{The NAK (No AcKnowledgement) message can be implemented as no-reply, if Alice has a time-out mechanism at her end.}. 

\item\label{step:check} When Alice receives $q^*$, which could be a valid preimage or a NAK, she checks the truth value of $H(q^*)=q^{[L-j]}$. If true, she stores\footnote{overwriting the previous value of $q$} $q^{[L-j-1]}=q^*$ and transitions to step \ref{step:alice}; the round is finished. {\bf If false}, she sends ($\delta_j$, $\tau_j$) again and remains in step \ref{step:check}.  
\end{enumerate}
\end{protocol}

It is easy to see that the protocol is robust. Alice could only be one step behind Bob if she has not received Bob's acknowledgement in round $j$, since Bob goes straight to round $j+1$ after he sends it. But then Alice will re-send her message, for which the acknowledgement is missing, and Bob will be able to see that the message is from the previous round and re-send his acknowledgement. By the channel model stated earlier after a finite number of retransmissions Alice will receive the correct acknowledgement and the parties will synchronise. We assume that if a round has exceeded the maximum number of retransmissions, the protocol fails due to DoS. 

Also note that step \ref{step:alice} has the highest communication cost as $\kappa$ hashes have to be communicated to Bob and in addition the document $\delta$. The document is typically hash-sized, since the actual document text can be communicated out of band, with only its digest being signed by the protocol. It would not be efficient to send $(\delta_j, \tau_j)$ again when, for example, only one element of $\tau_j$ is received with errors. However, there is an easy solution to this: send elements of $\tau_j$ one-by-one with immediate validation by Bob, who will hash them and compare the result to the stored top of the stack. This would limit retransmission to individual elements of $\tau_j$. When $\kappa$ of them have been received and confirmed, send $\delta_j$.  

\subsection {Security of the BWS protocol}
The security of the protocol rests on the following observations:
\begin{itemize}
\item Since the fabric is private to Alice, Lemma \ref{lem:stack-sec} applies, i.e. the combination of $E$ and $T$ uniquely defines $D$, making it impossible for Bob to forge Alice's signature for the current round. Hence Alice cannot repudiate a valid stack $(D,T)$ over a fabric with the edge $E$, when it is claimed by Bob.

\item However, Bob could claim to have received only some $S_{j_B}$, which is a substack $S_j$  corresponding to a smaller depth $j_B<j$. For sufficiently small $j_B$ Bob would be able to forge Alice's signature using fabric elements from $S_j$ and thus repudiate the genuine one. However, this scenario is still impossible, since, according to step \ref{step:check}, Alice holds the acknowledgement $a=q^{[L-j-1]}$ at the end of round $j$ and can prove that $\beta(a,q^{[L-1]})=j$. So Bob's stack cannot have less depth than $j$, which is the same value as Alice's. Consequently Bob cannot repudiate.   

\item Alice could try to repudiate differently. Since the fabric is available to her in its entirety, she could build 
a different stack $(D^\prime_j,T^\prime_j)$ of the same depth, not a substack of $(D,T)$, with generally different documents $D^\prime_j$. Alice could then claim that Bob has received and changed it. However, according to Lemma \ref{lem:stack-sec}, Bob could only have done it if he had access to Alice's private fabric (and collusion is outside the threat model of any bipartite signature protocol). Consequently, any dispute between Alice and Bob regarding any stack content should be resolved in favour of Bob automatically, provided that Bob's stack is valid.  
\end{itemize}

\subsection{Adjudication\label{sec:adj}}
BWS supports post-transaction adjudication by an algorithmic third party Judy, who need not authenticate Alice and Bob as long as she has the public values $E$ and $q^{[L-1]}$ authenticated out of band. Keeping to the non-collusion scenario, Judy performs the following steps. 
\begin{enumerate}
\item Judy requests from Alice the last $q=q^{[L-j-1]}$ she received from Bob, to determine the last $j$ in the process of its validation using the public $q^{[L-1]}$. Alice will not benefit from reducing $j$ to some $j_A<j$ since this would enable Bob to potentially forge her signature. However, Alice might chance it to falsely invalidate a few most recent rounds.
\item Judy requests Bob to provide the last $q$ he sent, which should be the same $q^{[L-j-1]}$. If in the process of validation the value of $j$ turns out to be some $j_B<j_A$, then Bob is lying and $j_A$ is accepted as the value of $d$, since Alice has no access to Bob's chain and since the only source of valid $q$ for her is Bob. If $j_B>j_A$, Judy accepts $j_B$ as the correct value of $d$. 
\item Judy requests a depth-$d$ stack over a fabric with the edge $E$ from Bob. Judy then validates the stack and confirms all signatures in it. 
\end{enumerate}

\subsection{Practicalities\label{sec:prac}}

\begin{figure}
\begin{center}
\includegraphics[width=1.0\textwidth]{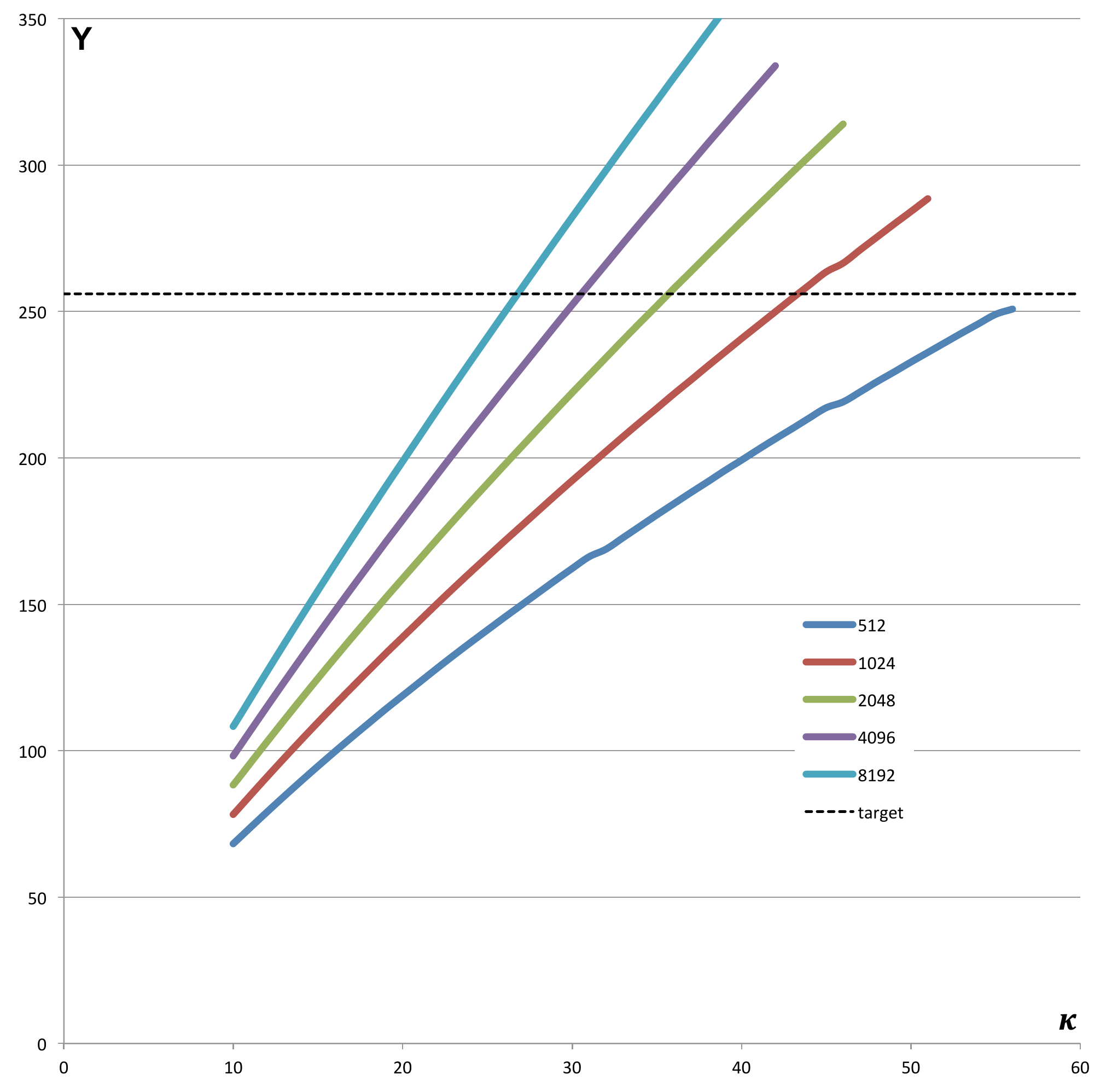}
\end{center}
\caption[~]{Security parameter ${\bf Y}$ vs oracle size $\kappa$ for various fabric widths. The dashed line marks the target security: 256 bits (or 128 bits Post Quantum). The target is reached at the following levels:\\
\begin{tabular}{|c|c||c|c||c|c||c|c||c|c|}
\hline
$w$&$\kappa$&$w$&$\kappa$&$w$&$\kappa$&$w$&$\kappa$&$w$&$\kappa$\\
\hline
512&--&
1024&44&
2048&36&
4096&31&
8192&27\\
\hline
\end{tabular}
}
\label{fig:oracle}
\end{figure}

Now let us discuss how practical an implementation of the BWS protocol can be. The issue boils down to analysis of three major cost parameters: storage, computation/power and communication. 

\paragraph{Storage.}
The issue of how much storage is required for the fabric is entangled with the issue of how much communication each round involves. Both are dependent on Eq \ref{eq:oracle} and the chosen security parameter ${\bf Y}=\log_2G$,  which is plotted against the oracle parameter $\kappa$ in Fig \ref{fig:oracle}. The curves are drawn up to a point at which $\kappa\log_2w\simeq512$. A further increase in $\kappa$ would necessitate a longer hash value than the output of SHA-512 for oracle emulation, which may be a problem. Also one has to remember that Eq \ref{eq:oracle} is only an approximation accurate for the region $\kappa\ll w$. For all the curves in Fig \ref{fig:oracle} $\kappa$ is at least one order of magnitude less than $w$, so the plots should be accurate enough. 

If we limit the discussion to the case when the security parameter is at least 256, we can see that it is impossible to reach that level with the fabric narrower than 512 if the standard hash SHA-512 is to be used. The width 512 is sufficient for a lesser, 192-bit, security but in the quantum case it would be reduced to one half, 89 bits, which may not be sufficient. The wider the fabric, the less the critical value of $\kappa$. According to Eq \ref{eq:capacity}, the required storage capacity to store the whole fabric is \[
\lambda w N \simeq \lambda\kappa d_{{\rm max}}\,,
\]
where $\lambda$ is the length of the hash used for the {\em chains} of the fabric (which is a parameter independent of the considerations of the oracle length). If the standard SHA-256 hash is used for the chains, $\lambda=32$ bytes. For example, for a fabric sufficient for $\sim$1M signatures with security ${\bf Y}=256$ it would appear that storage around 1G bytes would be required ($\kappa=31$, w=4096), which is a large but completely feasible amount. 

However, speed of access could  be traded off for the storage requirement to reduce it by orders of magnitude using hash recalculation. Indeed, instead of storing every member of the fabric $r^{[k]}_i$, choose a large positive $\phi$ and substitute $\phi k_1+k_2$, where  $k_2\in \llbracket0,\phi-1\rrbracket$, for $k$. Only store elements $r^{[k_1,k_2]}_i$ when $k_2=0$. When in the round part of the protocol and a value of some $r^{[k_1,k_2]}_i$ is required for $k_2\ne 0$, apply the hash function to $r^{[k_1,0]}_i$ $k_2$ times to obtain it. 

It is worth mentioning that the modern GPU-equipped PC's hashrate (to say nothing about a cluster's) is measured in billions per second, which makes it possible to compute any required fabric element $r^{[k]}_i$ from the initial random $r^{[0]}_i$ in a matter of milliseconds for any realistic fabric length, making it profitable to set $\phi=N$ and only store tens of kilobytes of the initial randoms. However, the protocols presented in this paper, due to their very low computation cost, may be an attractive option for embedded systems and the IoT as well. A typical cost of a hash calculation on a microcontroller is 10$\mu$s, down to 1$\mu$s with hardware acceleration. Expanding a single node of the Winternitz chain $r^{[0]}_i$ to the next $\phi$ nodes for $\phi\sim 1000$ would be a matter of single milliseconds. It would only be required once in 1000 reads from the fabric, which would amortise the computational cost nicely, while on the other hand it would reduce the storage requirements from gigabytes to much more affordable megabytes. 

Note that the cost of recalculation is on average {\em one} hash calculation per fabric element, since elements are recalculated in bunches of $\phi$ every $\phi$ steps up the chain and stored in a temporary buffer. There is a difference between the average computation requirements, which affect {\em energy} consumption and the application execution {\em time} on the one hand, and the peak computation load, which affects {\em latency}, {\em power} requirements and {\em cooling} on the other, so the trade-off between storage and recalculation in any particular case might be more subtle.  

All the above concerns Alice and Alice only. Bob has no access to the fabric. He only needs to store the documents and the top of the stack, which is the same size as Alice's public key $E$. The latter needn't be stored after the first document has been received, since on the one hand, $E$ is not needed for the round part of the protocol, which deals exclusively with the top of the stack, and on the other, it can be reconstructed at any round $j$ by popping documents off the stack in an obvious way (calculating the $\omega$ values of the documents starting from the last and working backwards). The need to produce $E$ may arise if Judy is involved. However, storing $E$ incurs only a small cost anyway, less than a 50\% increase in the required storage capacity, and is a diminishing share as more documents are received and need to be stored.

\paragraph{Communications.} The main consideration that drives the choice of the fabric width $w$ is the length of the fabric edge. Since the edge is used as the public key identifying the signer, the latter is interested in having it as short as possible, thus increasing the required $\kappa$ for a given security parameter. However, in step \ref{step:alice} of Protocol \ref{prot:bipartite} Alice sends $\kappa$ hashes (the difference between $T_j$ and $T_{j-1})$ in addition to the document (which is typically represented by the digest of the document file and is one hash in length). Consequently, there is a trade-off between the public key length and the signature length. Fig \ref{fig:oracle} indicates that the variation of $\kappa$ at our target level of security is rather limited, while the size of the public key doubles up every  time we widen the fabric. On the other hand, the public key is only communicated once, in step \ref{step:public-key}, while step  \ref{step:alice} of Protocol \ref{prot:bipartite} is invoked as many times as there are documents to be signed before the fabric is used up. This points to the largest affordable $w$ as the best solution. Increase in $w$ also helps to reduce the fabric length $N$ given the maximum depth $d_{max}$, which makes it possible to store fewer fabric elements for a given maximum recalculation cost. From this point of view, regimes close to $\kappa=31$, $w=4096$ seem optimal: 1K bytes to send in step \ref{step:alice} as a document digest and its signature, and 128K bytes to send in step \ref{step:public-key} as a public key, both easily within the capabilities of a low-bit-rate communication facilities available to an IoT device. 

\paragraph{Computations.} This is where the proposed protocol excels. Alice's costs for step \ref{step:alice} are trivial: one SHA-512 hash calculation as per Definition \ref{def:push} and a few table lookups to fetch the fabric elements, if they are 100\% stored. If they are recalculated, add $\kappa$ SHA-256 calculations as recalculation cost. One might think that the cost of the SHA-512 will increase as $j$ increases, since the new document is concatenated with all the previous ones thus making the hash argument ever longer, but this does not affect the cost. The mechanics of the hash algorithm are such that documents are processed block-by-block and the current state of the computation is used to produce the result. In round $j+1$ the hash computation will simply proceed from the point that it reached in round $j$ and will do the same fixed amount of work as that in round $j$. Another instance of hash calculation (SHA-256) is required at step \ref{step:check} to validate Bob's acknowledgement.  

Bob has a similar amount of work to do. At step \ref{step:bob}, Bob must validate Alice's message, which will cost $\kappa$ SHA-256 computations in any case, as it is not dependent on Alice's storage strategy; this is only a sub-millisecond time though, even for an IoT platform. There are also some table storage and retrieval operations to store the current top of the stack and retrieve the acknowledgement. If the acknowledgement chain is not 100\% stored and requires recalculation, add the cost of another SHA-256, but that is all Bob is spending on computations. 

\section{MAWS protocol\label{sec:MAWS}}

Even though the BWS protocol is a two-party transaction, the verifier party (Bob) can only verify that the document has been signed by the signing party (Alice). If Bob disagrees with the document itself, the only option he has is to refuse to acknowledge it, in which case Alice can only repeat the step either indefinitely or until the protocol detects denial of service. The only way out of the impasse is to reinitialise the protocol with a new fabric at a significant cost.

In this section we will present a solution which gives Bob the power to (in)validate the document at the same time as signing for its receipt. Such a solution is available immediately with the BWS protocol if two stacks are used, one for either party, with Alice and Bob swapping roles for the second stack. This way Alice signs a document using her stack as the signer, and Bob acknowledges as the verifier, then Bob signs his acceptance of the document using his stack as the signer, and Alice acknowledges the receipt of the acceptance as the verifier.  

However, it turns out that a single stack is sufficient to sign both the document and its acceptance. Under the bipartite protocol that we are about to present {\em both} parties are signers and both are verifiers, but one party has a significantly larger storage and communication (or, more precisely, transmission) requirements than the other. 

Before we define the protocol, let us simplify the rules somewhat. Instead of stating it explicitly, we will now assume that each message is validated by the receiver and if the validation fails, a NAK is sent back to the sender, but no change of state occurs at the receiver as if the message were never sent. Also we assume that either the channel is authenticated, in which case the NAK is an authenticated message, or the channel only has weak integrity, and then the NAK is in fact a time-out of a duration exceeding the time required for the maximum number of retransmissions. In both cases the receiving party will be able to identify a NAK with certainty, but in the latter case the reaction to a NAK should be the same as the one to an invalid message, since those can always be injected in the channel by a DoS attacker over large enough period of time. Since we only require weak integrity, the protocols in the sequel will not differentiate between NAKs and invalid messages. 

The protocol is fully asynchronous, i.e. each send requires a valid acknowledgement to be received. In the absence of an acknowledgement, the sending party re-sends its message up to the retransmission limit, then the protocol fails. The protocol does not require the channel between Alice and Bob to have absolute integrity; as before, we mark received values with an asterisk $*$ to emphasise that they are not necessarily the same as those sent.  

\begin{protocol} [Mutual Asymmetric Winternitz Stack (MAWS) Protocol]\label{prot:maws}
~\\
Parties: Alice and Bob\\
Protocol parameter: fabric width $w\in\mathbb{N}$, $w$ is a power of 2.\\
{\bf Initially:} as in Protocol \ref{prot:bipartite}

\vskip1em
\noindent
{\bf Repeat for $j=2,4..L-1$ [only even numbers]:}

\begin{enumerate}[label={\bf R}\rm\arabic*]
\item \label{step:alice*} A new transactions document $\delta_j$ requires signing. Alice computes \[
S_{j}={\rm\bf p}(\delta_j,S_{j-1})=(D_j,T_j)\,,
\]
\noindent stores it in local memory overwriting $S_{j-1}$ and computes $\tau_j=T_j - T_{j-1}$. 

\item \label{step:alice**}
Alice sends $(\delta_j,\tau_j)$ to Bob and goes to step \ref{step:check**} to await an acknowledgement.

\item \label{step:bob*}Bob receives $(\delta^*_j,\tau^*_j)$ and validates it by $\epsilon(\delta_j^*,\tau^*_j,S_{j-1})$. 
If valid, Bob concludes that
\[\delta_j=\delta_j^*\;\; \hbox{and}\;\; \tau_j=\tau_j^*\,,
\] 
and stores 
\[
S_j=(D_{j-1}\triangleright\delta_j,T_{j-1}+\tau_j)\,,
\]
where $(D_{j-1},T_{j-1})=S_{j-1}$, in local memory overwriting $S_{j-1}$. If Bob approves $\delta_j$, he forms his signature \[
\delta^\prime_j=H\left(\delta_j \concat q^{[L-j-2]}\right)\,,
\] 
otherwise he sets $\delta^\prime_j$ to zero.
\item \label{step:bob**}
Bob sends the pair $(q^{[L-j-1]},\delta^\prime_j)$ as the acknowledgement to Alice and goes to step \ref{step:ack} to await an acknowledgement. 

\item\label{step:check**} Alice retracts to step \ref{step:alice**} unless she receives $(q^{[L-j-1]*},\delta^{\prime*}_j)$ validated by $H(q^*)=q^{[L-j]}$. If continuing, Alice stores $q^{[L-j-1]}=q^{[L-j-1]*}$, computes  \[
S_{j+1}={\rm\bf p}(\delta^{\prime*}_j,S_{j})=(D_{j+1},T_{j+1})\,,
\]
\noindent and stores it in local memory overwriting $S_j$, while computing $\tau_{j+1}=T_{j+1} - T_j$.

\item\label{step:check***} 
Alice sends the pair $(\delta^{\prime*}_j,\tau_{j+1})$ to Bob and waits for acknowledgement at step \ref{step:final}

\item\label{step:ack} Bob retracts to step \ref{step:bob**} unless he receives $(\delta^{\prime**}_j,\tau_{j+1}^*)$, validated by $\epsilon(\delta^{\prime**}_j,\tau_{j+1},S_j)$. If valid, Bob concludes that \[
\delta^{\prime**}=\delta^{\prime*}_j\;\;\hbox{and}\;\;\tau_{j+1}^*=\tau_{j+1}\,.
\]
and stores
\[
S_{j+1}=(D_j\triangleright\delta^{\prime*}_j,T_j+\tau_{j+1})\,,
\] 
where $(D_j,T_j)=S_j$, overwriting $S_j$,

\item\label{step:ack*}
Bob sends $q^{[L-j-2]}$ as an acknowledgement to Alice. If $\delta^{\prime*}_j=\delta^{\prime}_j$, the transactions is completed, and Alice knows it. Otherwise, Bob's signature $\delta^\prime_j$ was either zero or miscommunicated, and again, Alice knows it. The transaction is null and void; a (complete) repeat-round is necessary if both parties still wish to sign.

\item\label{step:final} Alice retracts to step \ref{step:check***} unless she  receives $q^{[L-j-2]*}$ and validates it by $H(q^{[L-j-2]*})=q^{[L-j-1]}$. If valid, she stores $q^{[L-j-2]}=q^{[L-j-2]*}$ and checks that 
\[
\delta^{\prime*}_j=H\left(\delta \concat q^{[L-j-2]}\right)\,.
\] 
If the equation holds, then the transaction is completed and Bob knows it. Otherwise Bob either rejected the transaction or his approval was miscommunicated, Bob knows which. Either way, the transaction is null and void; a (complete) repeat-round is necessary if both parties still wish to sign.
\end{enumerate}
\end{protocol}

The protocol is generally robust as the sending of a message is paired with its validation and possible retransmission, except step \ref{step:bob**}, where Bob's signature is communicated but it cannot be validated  {\em before} step \ref{step:ack} when Alice has already sent $\kappa$ hashes and Bob's signature back. If the signature was corrupted in communication at step \ref{step:ack}, this would waste the protocol round, in terms of both communication and fabric/chain material. 

\paragraph{Security of the MAWS protocol.} The non-repudiation properties of MAWS hinge on the fact that Bob's signature is based on the pre-image of the latest member of Bob's Winternitz chain disclosed to Alice, namely $q^{[N-j-2]}$. Alice is unable to forge Bob's signature without knowledge of $q^{[N-j-2]}$, and when that value is disclosed to her in step \ref{step:ack*}, Bob's signature or refusal to sign has been signed by Alice already, in steps \ref{step:check**} and \ref{step:check***}. 

So it looks as though without hosting a Winternitz fabric, Bob can sign Alice's documents using nothing more than a single hash chain. The post-transaction adjudication for MAWS is the same as that for BWS, see section \ref{sec:adj}, except Judy also checks all $\delta^\prime$ messages and marks the documents as approved or not approved accordingly.

\section{Reverse Winternitz Stack (RWS) Protocol\label{sec:RWS}}

Now let us tighten the communication model. Since the DoS defences would benefit from filtering incoming messages before the protocol calculations based on them are launched anyway, let us assume that Alice and Bob share a weak secret and use a symmetric Message Authentication Code (MAC, e.g. HMAC, based on the same hash function as the chains) to authenticate messages from Alice to Bob and back. Message authentication cannot be used to replace signatures since MACs are symmetric and can be repudiated. However, even a short MAC stops message insertion and message altering attacks very effectively. For example, a 32-bit MAC has less than one in a billion chance to be guessed in an attempt to insert or alter a message. On the other hand, for a known-plaintext attack to succeed in obtaining even a short AES128 key, the number of intercepted messages required is many orders of magnitude more than the size of any realistic Winternitz fabric. 

Let us therefore adopt a more restrictive channel model, whereby a message sent is extremely likely to be received correctly or not at all. Under such conditions MAWS becomes robust and any validation failure can safely be attributed to protocol violation by the sending party. 

Given that we are now able to propose a protocol where Alice has no independent signing function. All Alice does is certify Bob's signatures, effectively turning into a kind of secure signature server. Alice is unable to forge Bob's signature, nor Bob repudiate it. Assuming non-collusion, mutually mistrustful Alice and Bob are still able to prove to Judy that Bob signed the documents he claims to have signed and to stop him repudiating his signature. The security of the following protocol trivially follows from the security of MAWS.

\begin{protocol} [Reverse Winternitz Stack (RWS) Protocol]\label{prot:reverse}
~\\
Parties: Alice(signature server) and Bob(signer)\\
Protocol parameter: fabric width $w\in\mathbb{N}$, $w$ is a power of 2.\\
{\bf Initially:} as in Protocol \ref{prot:bipartite}
\vskip1em
\noindent
{\bf Repeat for $j=2..L-1$:}

\begin{enumerate}[label={\bf R}\rm\arabic*]
\item\label{step:xbob} Bob computes signature \[
s_j=H\left(\delta_j \concat q^{[L-j-2]}\right)\,,
\] 
where $\delta_j$ is the document he wishes to sign (or its hash, whichever is shorter), and sends it to Alice via an authenticated channel. 

\item\label{step:xalice} Alice eventually receives $s_j$ from Bob and 
computes \[
S_{j}={\rm\bf p}(s_j,S_{j-1})=(D_j,T_j)\,,
\]
\noindent stores $S_j$ and sends to Bob $\tau_j=T_j-T_{j-2}$.

\item\label{step:xbob*} Bob eventually receives $\tau_j$ and validates it by $\epsilon(s_j,\tau_j,S_j-1)$. If invalid, the protocol fails. Otherwise,
Bob stores
\[
S_j=(D_{j-1}\triangleright s_j,T_{j-1}+\tau_j)\,,
\] 
where $(D_{j-1},T_{j-1})=S_{j-1}$, overwriting $S_{j-1}$. Bob also stores $\delta_j$ under the index $j$ and sends $q^{L-j-2}$ to Alice as the acknowledgement. 

\item\label{step:xfinal} Alice eventually receives $q^{[L-j-2]}$, checks that $H(q^{[L-j-2]})=q^{[L-j-1]}$ and if so, stores $q^{[L-j-2]}$ overwriting $q^{[L-j-1]}$ and completes the round. Otherwise the protocol fails.
\end{enumerate}
\end{protocol}

The protocol only fails if a party wilfully sends the wrong message. Failure to receive a response should be construed as a communication failure, not a security event, as the protocol stalls awaiting retransmission.

As before, a repudiation attempt from Bob on a given document $\delta$ will be countered by the retrieval of the latest known $q$ from Alice and a stack of the corresponding depth from Bob. The verifier will then examine all $s_j$ to find the one for which $s_j=H\left(\delta \concat q^{[L-j-2]}\right)$, which proves the signature. 

The last of the signed documents introduces an uncertainty as to whether or not Alice has completed step \ref{step:xfinal} on it or not, but it is not a major problem. The verifier may simply delay verification until Alice is quiescent.

\section{Application of RWS to Internet of Things\label{sec:app}}

The variety of IoT devices is very broad. It stretches from systems that have computation and communication capabilities approaching those of ordinary computers, to microcontroller-based smart sensors on a tight energy budget with low-bit-rate long-range (LoRa) radio communications. It is the latter category that present unique challenges in network security, especially when nonrepudiation is required in a multivendor safety-critical system. 

It is little appreciated in literature that IoT communication requirements are quite asymmetric. The success of sending  data over the radio depends very much on the transmit power, which has to come out of the overall power budget of the device. IoT platforms tend to transmit little and do it infrequently. There are also legal constraints on the duty cycle and radiated power when operating in the frequency bands available to LoRa transmissions. However, receiving data is possible at a higher data rate spending much less power. In fact the power is used mostly for digital signal processing of the received signals, not the reception process as such; it can be reduced further by doing the processing less fast. In an IoT swarm, the edge server is typically equipped with a more powerful transmitter operating at a higher bit-rate. The IoT device can receive such a signal with less battery drain. 

Ordinary non-repudiation protocols, e.g. producing the elliptic curve signatures of a given document by both parties to a transaction, exhibit symmetric computation and communication requirements. For example a 256-bit ECDSA produces a 512-bit or 64 byte signature, which is not quantum secure. Moreover, even using advanced microprocessors (rather than cheap microcontrollers), such as ARM Cortex-M4, the signature computation time is measured in hundreds of milliseconds compared to hundreds of microseconds\cite{ECDSA} for $\sim$30 hashes that Bob computes in each round of RWS ($w=4096$). The volume of data transmitted by Bob in one round of RWS is the same as it is under ECDSA, namely 32 bytes for the signature $s$ and 32 bytes for the acknowledgment $q$. The volume of data to receive for Bob is much higher, close to 1 Kbyte. Finally, Bob retains the audit trail of all his signatures with the assurance that all of them (possibly with the exception of the very last one) have been registered by Alice and hence usable in transactions.

The audit trail forces Alice to be honest, especially when there is a system penalty if a proof against Alice is submitted by Bob. If Alice knows that, and if there exists some Proof of Stake for Alice (not necessarily digital), she can be used as Bob's proxy, making it unnecessary for a third party to communicate with Bob for signature validation frequently, especially if delayed validation is compatible with the security model. e.g. in the airplane black box type of application. 

\section{Related work\label{sec:rel}}
The idea of a hash-based signature scheme is classic, due to Lamport \cite{OTS}. The basic approach is to associate a pair of nonces with each digit (one for the value 0 and the other for the value 1) of a message digest and use their hashes as the public key. The signer reveals the nonce associated with the value of the corresponding digit to form the message signature, which is only effective for one message. The scheme involves communication of very large signatures. This was improved upon by Merkle \cite{M-OTS} and Winternitz. The former proposal is to only sign the digits where the value is 1 and to include a checksum to counter bit omission. The latter proposal (Winternitz One-Time Signature, WOTS) is to segment the digest into chunks and use each chunk as an iteration counter in repeatedly hashing the corresponding nonce, again with a checksum guarding against reduction of iteration counters. WOTS was first published as an idea outline in Merkle's conference paper \cite{MerkWin}, reference 6 of which is to Winternitz's private communication. Neither Merkel nor Winternitz proposed anything  to mitigate the one-time nature of Lamport's signature protocol, and the improvements are only in the signature size, which is much shorter than Lamport's, but is still very long compared to public-key cryptography with similar security parameters. This line of research has been continued further; more recent work includes a WOTS$+$\cite{WOTS+} scheme, which extends WOTS, and XMSS\cite{XMSS}, which extends the original Merkle proposal. 

Another line of research was sprung by the seminal paper on HORS\cite{HORS}, a "Hash to Obtain Random Subset" proposal, which turned out to be very fruitful. The idea here is to hash the digest and partition the hash image into equal bit-size unsigned integers, the values of which are gathered into an index set. The indices select the pre-images to be revealed to form the message signature. An attacker would have to find a different message whose digest produces a subset of the index set under the hash function to obtain a false signature. The authors of \cite{HORS} demonstrate that this is computationally hard, and the hardness remains significant even when the set of pre-images used has expanded after signing a few messages. This approach and its successors are often referred to as {\em few-time} signature schemes. There is an elaboration of HORS by H{\"u}lsing in \cite{SPHINCS}, where the ideas of WOTS$+$ and an improved version of HORS, HORST, are combined. For the security parameter value 256, which we use as the typical case, the message signature claimed in \cite{SPHINCS} is 41 thousand bytes, see Table 1 in that paper. This does not compare favourably with about one thousand bytes in the MAWS signature in the typical case, and even less with 64 bytes in RWS (all cases, ignoring input volume). However, the upside of their approach is a short public key, circa 1K bytes, whereas our method would require a public key measured in hundreds of kilobytes (128K in the typical case). However the public key can be stored in an embedded system or obtained by its hash via an unprotected public network, so we do not see the size of the public key as an important parameter.  

HORS-like signatures can be shortened further: a recent paper, \cite{HORSIC} claims a reduction by more than a third, but the ballpark cost of communicating a signature of this size is still more than an order of magnitude more expensive than any message of our protocols. Finally, we are aware that the known few-time signatures have striven to rid themselves of the protocol state as this is seen as undesirable in the general security setting, see \cite{SPHINCS}. Our protocols are clearly not stateless; however, for the application domain they are intended for it can be an advantage, since not only the transactions but also their ordering is assured by the signature stack; neither can be repudiated.

\section*{Conclusions}

The Winternitz stack over a fabric has been proposed as a basis of post-quantum digital signature protocols. The security properties of the stack have been studied and a few protocols derived from them. The most interesting protocol, RWS, has a short signature size, 64 bytes, yet it is exclusively hash-based and ensures nonrepudiation. The security level of the protocol depends solely on Alice's available communication resources and Bob's storage constraints, but not on the parties' computational resources. For a reasonable amount of storage and compute power the protocol achieves 256-bit classical security or 128-bit quantum one, which does not diminish as more messages are signed without refreshing the public key. Under realistic assumptions at least 1 mln signatures can be made under the same public key, more if storage/recomputation is not a problem. 

The protocol places communication requirements on Alice and Bob asymmetrically, with Alice mostly transmitting and Bob receiving, which helps with the constraints of the low bit-rate communication characteristic of the IoT (in particular sensor networks). Alice can then act as a signature server and Bob a client without requiring any trust between them. The protocol provides sufficient assurances to Alice that Bob's signature is genuine, but if this needs to be proven to a third-party adjudicator {\it post hoc}, Alice and Bob must each supply their respective validation data. Given a no-collusion threat model, which is mandatory for all bipartite protocols, Bob cannot repudiate. Furthermore Bob cannot sign a document without Alice being aware of it and Alice is unable to forge Bob's signature to the adjudicator since Bob's signatures come from Bob as part of his validation data.

The author is sincerely grateful to Bruce Christianson who has read the manuscript and provided very useful feedback. 
  
\bibliographystyle{plainurl}
\bibliography{wss}

\begin{thebibliography}{10}

\bibitem{SPHINCS}
Daniel~J. Bernstein, Daira Hopwood, Andreas H{\"u}lsing, Tanja Lange, Ruben
  Niederhagen, Louiza Papachristodoulou, Michael Schneider, Peter Schwabe, and
  Zooko Wilcox-O'Hearn.
\newblock {SPHINCS}: Practical stateless hash-based signatures.
\newblock In Elisabeth Oswald and Marc Fischlin, editors, {\em Advances in
  Cryptology -- EUROCRYPT 2015}, pages 368--397, Berlin, Heidelberg, 2015.
  Springer Berlin Heidelberg.

\bibitem{binomials}
Shagnik Das.
\newblock A brief note on estimates of binomial coefficients.
\newblock 2015.
\newblock URL: \url{http://page.mi.fu-berlin.de/shagnik/notes/binomials.pdf}.

\bibitem{ECDSA}
Hayato Fujii and Diego~F. Aranha.
\newblock {Curve25519} for the {Cortex-M4} and beyond.
\newblock In {\em Progress in Cryptology {\textendash} {LATINCRYPT} 2017},
  pages 109--127. Springer International Publishing, 2019.

\bibitem{WOTS+}
Andreas H{\"{u}}lsing.
\newblock {WOTS$+$} -- shorter signatures for hash-based signature schemes.
\newblock Cryptology ePrint Archive, Report 2017/965, 2017.

\bibitem{XMSS}
Andreas H{\"{u}}lsing, Denis Butin, Stefan{-}Lukas Gazdag, Joost Rijneveld, and
  Aziz Mohaisen.
\newblock {XMSS:} {eXtended} {M}erkle signature scheme.
\newblock {\em {RFC}}, 8391:1--74, 2018.

\bibitem{OTS}
Leslie Lamport.
\newblock {Constructing digital signatures from a one-way function. Vol. 238.
  Technical Report CSL-98}.
\newblock Technical report, SRI International, 1979.

\bibitem{HORSIC}
Jaeheung Lee and Yongsu Park.
\newblock {HORSIC}$+$: An efficient post-quantum few-time signature scheme.
\newblock {\em Applied Sciences}, 11(16):7350, aug 2021.

\bibitem{M-OTS}
Ralph~C. Merkle.
\newblock {\em Secrecy, Authentication and public-key cryptosystems}.
\newblock UMI Research Press, 1982.

\bibitem{MerkWin}
Ralph~C. Merkle.
\newblock A digital signature based on a conventional encryption function.
\newblock In Carl Pomerance, editor, {\em Advances in Cryptology --- CRYPTO
  '87}, pages 369--378, Berlin, Heidelberg, 1988. Springer Berlin Heidelberg.

\bibitem{HORS}
Leonid Reyzin and Natan Reyzin.
\newblock Better than {BiBa}: Short one-time signatures with fast signing and
  verifying.
\newblock In Lynn Batten and Jennifer Seberry, editors, {\em Information
  Security and Privacy}, pages 144--153, Berlin, Heidelberg, 2002. Springer
  Berlin Heidelberg.

\end{thebibliography}

\end{document}